\documentclass{ws-procs975x65}

\def\rmd{\mathrm{d}}

\begin{document}

\title{Gravitational lensing in area metric spacetimes}
\author{Marcus C. Werner}

\address{Center for Gravitational Physics, Yukawa Institute for Theoretical Physics, \\ Hakubi Center for Advanced Research, \\ Kyoto University,\\
Kitashirakawa Oiwakecho Sakyoku, Kyoto 606-8502, Japan \\
E-mail: werner@yukawa.kyoto-u.ac.jp}

\begin{abstract}
We consider light propagation as a probe of non-metricity in area metric spacetimes, and find a deviation from the standard Etherington relation for linearized area metric Schwarzschild. This is joint work with Frederic P. Schuller (Erlangen University).
\end{abstract}


\bodymatter

\section{Introduction}
Since the 1919 eclipse expeditions, whose centennial will be celebrated in the year following this Marcel Grossmann Meeting, gravitational lensing has provided important precision tests of the Lorentzian metric spacetimes of general relativity. This raises the question of how modified theories with {\it non-metric} spacetime structures may be probed by gravitational lensing.

The answer to this question requires, of course, knowledge of the gravitational theory of such non-metric spacetime kinematics, and the standard approach is to start by stipulating some modification of the Einstein-Hilbert action as the gravitational dynamics. However, recent work in geometrodynamics (cf. Ref.~\refcite{gsww12}, \refcite{dssw18}) has shown that predictive gravitational dynamics can, in fact, be {\it derived} from the underlying spacetime kinematics, such that the theory is predictive by construction. Hence, this new approach is called {\it constructive gravity}, and session AT5 of this meeting is dedicated to it.

Here, we describe the first concrete example of how constructive gravity can be employed to derive a prediction for gravitational lensing in a non-metric spacetime. Specifically, we consider area metric geometry for which a perturbative Schwarzschild-like solution has been obtained, and find a deviation from the standard metric Etherington distance duality relation. This proceedings paper is based on Ref.~\refcite{sw17} and references therein.

\section{Geometrical background}
\subsection{General kinematics}
\label{sec:kinematics}
We begin by discussing general spacetime kinematics and the notion of predictivity, following Ref.~\refcite{rrs11}. Consider a smooth 4-dimensional manifold $M$ endowed with smooth tensor fields $G$ and $F$, which we refer to as {\it geometry} and its {\it test matter}, respectively, governed by a general linear field PDE,
\begin{equation}
\left[\sum_{d=1}^k D^{\bar{\mu}\nu_1 \ldots \nu_d}_{\bar{\lambda}}(G) \frac{\partial}{\partial x^{\nu_1}}\ldots \frac{\partial}{\partial x^{\nu_d}}  \right] F_{\bar{\mu}}=0\,,
\label{pde}
\end{equation}
where $\nu_i \in \{0,\ldots, 3\}$ denote general spacetime coordinates, $\bar{\mu}$ multi-indices of test matter fields components, and $i\in \{1,\ldots, d\}$, \ $d\in \{1, \ldots, k\}$ partial derivative order where $k$ is highest. Now, causal properties of Eq. (\ref{pde}) are governed by its principal polynomial, which emerges from the eikonal approximation: letting
\[
F_{\bar{\mu}}(x,\epsilon)=e^{\frac{{\rm i}S(x)}{\epsilon}}\sum_{j=0}^\infty F_{\bar{\mu}j}(x)\epsilon^j\,, \quad \mbox{and} \quad \epsilon\rightarrow 0\,,
\]
where $S$ is the eikonal (phase) function, then from (\ref{pde}),
\[
e^{\frac{{\rm i}S(x)}{\epsilon}}\left(\frac{{\rm i}}{\epsilon}\right)^{k}\left[D^{\bar{\mu}\nu_1\ldots \nu_k}_{\bar{\lambda}}(x) \frac{\partial S}{\partial x^{\nu_1}}\ldots\frac{\partial S}{\partial x^{\nu_{k}}}\right]F_{\bar{\mu}0}(x)+ \ \mbox{lower terms in}\ \frac{1}{\epsilon}=0\,,
\]
whence the first term stemming from the highest derivative order remains in the limit, and for this equation to have non-trivial solutions, the determinant of the square bracketed term needs to vanish. This yields the principal polynomial of Eq. (\ref{pde}) $P: T^\ast M\rightarrow \mathbb{R}$, reducing repeated powers,
\begin{equation}
P(x,p)=w_G \det  \left[D^{\bar{\mu}\nu_1\ldots \nu_{k}}_{\bar{\lambda}}(x) p_{\nu_1}\ldots p_{\nu_{k}}\right]=P^{\nu_1\ldots \nu_{\deg P}}p_{\nu_1}\ldots p_{\nu_{\deg P}}\,,
\label{pp}
\end{equation}
where $P^{\nu_1\ldots \nu_{\deg P}}$ is the totally symmetric principal polynomial tensor, and $w_G$ some appropriate weight function to render $P$ a scalar. Thus, the general {\it null cone} (or, null dispersion relation) at $x \in M$ becomes
\begin{equation}
N_x=\{p \in T_x^\ast M\ : \ P(x,p)=0\}\,.
\label{null}
\end{equation}

We are interested in causal kinematics of the generalized spacetime $(M,G,F)$, which is determined by the Cauchy problem. Given (\ref{pde}) and initial data, the Cauchy problem is well-posed if there is a unique solution depending continuously on the initial data, so $P$ is necessarily hyperbolic.

So far, we have only considered covectors, or particle momenta. However, we also need the dual vectors for particle trajectories. Now it turns out that if $P$ is hyperbolic, then its dual polynomial $P^\sharp: TM\rightarrow \mathbb{R}$ exists, although hyperbolicity of $P$ does not imply hyperbolicity of $P^\sharp$. But for predictivity in the sense of distinguishing future and past, or the sign of particle energies, $P^\sharp$ needs to be hyperbolic as well. This is called {\it bihyperbolicity}, and is the geometric requirement for predictive kinematics.

{\bf Example: standard Maxwell.} In $M=\mathbb{R}^4, \ x^\nu=(t,\mathbf{x})$, consider Maxwell's equations in vacuum, in suitable units. Then we have two constraint equations $\nabla\cdot \mathbf{E}=0, \ \nabla \cdot \mathbf{B}=0$, and two dynamical equations,
\[
\frac{\partial \mathbf{E}}{\partial t}-\nabla \times \mathbf{B}=0\,, \qquad
\frac{\partial \mathbf{B}}{\partial t}+\nabla \times \mathbf{E}=0\,.
\] 
Introducing $F_{\bar{\mu}}=(-\mathbf{E},\mathbf{B})$, $\bar{\mu}\in \{1,\ldots, 6\}$, they can be recast as test matter field equations in the form of Eq. (\ref{pde}) to obtain
\[
D^{\bar{\mu}\nu}_{\bar{\lambda}}\frac{\partial F_{\bar{\mu}}}{\partial x^\nu}=0\,.
\]
Now according to Eq. (\ref{pp}), the corresponding principal polynomial is computed from
\[
\det [ D^{\bar{\mu}\nu}_{\bar{\lambda}}p_\nu]=\det \left[
\begin{array}{*{6}c}
-p_0 & 0 & 0 & 0 & p_3 & -p_2 \\
0 & -p_0 & 0 & -p_3 & 0 & p_1 \\
0 & 0 & -p_0 & p_2 & -p_1 & 0 \\
0 & p_3 & -p_2 & p_0 & 0 & 0 \\
-p_3 & 0 & p_1 & 0 & p_0 & 0 \\
p_2 & -p_1 & 0 & 0 & 0 & p_0
\end{array}
\right]=p_0^2(-p_0^2+p_1^2+p_2^2+p_3^2)^2\,,
\]
whence, after reducing repeated powers, we read off the principal polynomial,
\[
P(x,p)=-p_0^2+p_1^2+p_2^2+p_3^2=P^{\mu\nu}p_\mu p_\nu=\eta^{\mu\nu}p_\mu p_\nu\,,
\]
identifying the principal polynomial tensor as the inverse Minkowski metric.

\subsection{Area metric spacetimes}
We can now specialize the previous discussion to test matter defined by general linear electromagnetism, which has the Lagrangian 
\begin{equation}
L=-\frac{1}{8}\mathcal{G}^{\mu\nu\rho\sigma} F_{\mu\nu}F_{\rho\sigma}\,, 
\label{gle}
\end{equation}
where $F_{\mu\nu}=-F_{\nu\mu}$ is the electromagnetic tensor, and $\mathcal{G}$ is called constitutive tensor density, describing the properties of the optical medium. This is the structure defining pre-metric electromagnetism (see, e.g., Ref. \refcite{ho03}), although in the standard Maxwell vacuum we consider $\mathcal{G}$ to be induced by a Lorentzian spacetime metric $g$. However, more generally, $\mathcal{G}$ is an area metric structure.

An {\it area metric} is a smooth 4th order tensor field $G$ with symmetries such that
\[
G_{\mu\nu\rho\sigma}=G_{\rho\sigma\mu\nu}\,, \ G_{\mu\nu\rho\sigma}=-G_{\nu\mu\rho\sigma}\,, \ G_{\mu\nu\rho\sigma}=-G_{\mu\nu\sigma\rho}\,,
\]
and we define an area metric induced by a metric $g$ as follows,
\[
(G_g)_{\mu\nu\rho\sigma}=g_{\mu\rho}g_{\nu\sigma}-g_{\mu\sigma}g_{\nu\rho}-\sqrt{\det g}\ \epsilon_{\mu\nu\rho\sigma}\,.
\]
The name stems from its area measuring property\footnote{To see this, consider two vectors $X,Y$ and a Riemannian metric $g$, then 
\[
G_g(X,Y,X,Y)=(X\cdot X)(Y\cdot Y)-(X\cdot Y)^2=|X|^2|Y|^2\sin^2\angle(X,Y)=|X\wedge Y|^2\,,
\]
that is, the squared area spanned by $X,Y$.}. Now {\it vacuum} electromagnetism on an area metric background can be described like Eq. (\ref{gle}) with $F$ subject to
\begin{equation}
L=-\frac{1}{8}\mathcal{G}^{\mu\nu\rho\sigma} F_{\mu\nu}F_{\rho\sigma} \quad \mbox{with} \quad \mathcal{G}^{\mu\nu\rho\sigma}=\omega_G G^{\mu\nu\rho\sigma}\,,
\label{amel}
\end{equation}
where $\omega_G^{-1}=\frac{1}{4!}\epsilon_{\mu\nu\rho\sigma}G^{\mu\nu\rho\sigma}$. Thus, an area metric spacetime with predictive kinematics in the sense of Sec. \ref{sec:kinematics} can be defined as the triple $(M,G,F)$ with a corresponding bihyperbolic principal polynomial $P$.

\section{Gravitational lensing}
\subsection{Light propagation}
\label{sec:light}
Let us know proceed to study light propagation on such an area metric spacetime $(M,G,F)$. As discussed in general in Sec. \ref{sec:kinematics}, we can apply the eikonal approximation to the field equations of Eq. (\ref{amel}) to define geometrical optics. The null cone (\ref{null}) in this case is, of course, found to be {\it quartic} ($\deg P=4$),
\[
P(x, p)=P(G)^{\alpha\beta\gamma\delta}p_\alpha p_\beta p_\gamma p_\delta=0\,,
\]
with the principal polynomial tensor constructed from the area metric $G$. It is clear, then, that light rays in such a spacetime are, in general, subject to {\it birefringence}.

However, since observationally any such effect would necessarily be very small, we shall from now on investigate a perturbation about an effective Minkowski spacetime. Then the (inverse) area metric is given by
\[
G^{\mu\nu\rho\sigma}=\eta^{\mu\rho}\eta^{\nu\sigma}-\eta^{\mu\sigma}\eta^{\nu\rho}-\epsilon^{\mu\nu\rho\sigma}+H^{\mu\nu\rho\sigma}\,,
\]
where $H$ is small. In this case, the principal polynomial tensor and its dual are found to be merely {\it quadratic},
\begin{equation}
P^{\mu\nu}=\eta^{\mu\nu}+H^{\mu\nu}\,, \quad  P^\sharp_{\mu\nu}=\eta_{\mu\nu}-H_{\mu\nu}\,,
\label{p}
\end{equation}
where $H^{\mu\nu}=\frac{1}{2}H^{\alpha\mu\beta\nu}\eta_{\alpha\beta}-\frac{1}{4!}\epsilon_{\alpha\beta\gamma\delta}H^{\alpha\beta\gamma\delta}\eta^{\mu\nu}$. Hence, there is an {\it effective} Lorentizan metric $P^\sharp_{\mu\nu}$ for light rays, and standard ray optical arguments for light rays apply. This fact, however, does not eliminate effects of the non-metricity of this spacetimes entirely: one result important for our later discussion is that energy-momentum conservation (for the general case, cf. Ref. \refcite{gim04}) implies that the photon current $N^\mu$ obeys the conservation law
\begin{equation}
(\omega_G N^\mu)_{,\mu}=0\,,
\label{conservation}
\end{equation}
which corresponds to the well-known result for metric spacetimes (see, e.g., Eq. 3.26 of Ref. \refcite{sef92}), except that the volume measure here, $\omega_G$, is area metrical.

\subsection{Linearized gravity}
\label{sec:gravity}
Having established our non-metric spacetime kinematics, we need gravitational dynamics next in order to obtain a concrete prediction for light propagation. But rather than stipulating a modification of general relativity, contructive gravity can be employed to derive it.

This geometrodynamical technique originates in Ref.~\refcite{hkt76}, which treats the Lorentzian metric case, but has been extended recently to any tensorial background (see Ref.s~\refcite{gsww12} and \refcite{dssw18}). Briefly, the idea is that a bihyperbolic spacetime kinematics yields a generalized ADM split with lapse and shift, defining a hypersurface deformation algebra. Recognizing that kinematical deformation and dynamical evolution coincide, the constraint algebra with supermomentum and superhamiltonian can be determined, resulting eventually in a PDE system for the gravitational Lagrangian.

The gravitational dynamics for area metric kinematics can, so far, be derived perturbatively. The solution for a point mass $M$, that is, the linearized area metric Schwarzschild, is
\begin{align*}
\hspace{-1cm}
G^{0a0b}\equiv&-\gamma^{ab}+H^{0a0b} = -\gamma^{ab}+(2A-\tfrac{1}{2} U + \tfrac{1}{2}V)\gamma^{ab}\,,  \\
G^{0bcd}\equiv& \epsilon^{bcd}+H^{0bcd} = \epsilon^{bcd}+\left(\tfrac{3}{4}U-\tfrac{3}{4}V-A\right)\epsilon^{bcd}\,,\\
G^{abcd}\equiv& \gamma^{ac}\gamma^{bd}-\gamma^{ad}\gamma^{bc}+H^{abcd} = (1+2U-V)(\gamma^{ac}\gamma^{bd}-\gamma^{ad}\gamma^{bc})\,,
\end{align*}
with Euclidean metric $\gamma^{ab}$, Euclidean distance $r$ of the unperturbed background, Levi-Civita symbol $\epsilon^{abc}$, and scalar perturbations
\[
A = - \frac{M}{8\pi r}\left(\kappa - \lambda \eta e^{-\mu r}\right)\,, \
U = - \frac{M}{4\pi r} \eta e^{-\mu r}\,, \
V = \frac{M}{4\pi r}\left(\kappa - \tau\eta e^{-\mu r}\right)\,,
\]
where $\eta, \kappa, \lambda, \mu, \tau$ are constants.

\subsection{Modified Etherington relation}
Finally, we are ready to study light propagation in this non-metric spacetime under the influence of gravity. One fundamental kinematical property of {\it any} Lorentzian metric spacetime is the Etherington distance duality relation, originally derived in Ref.~\refcite{e33}, which connects the luminosity distance, angular diameter distance and redshift. Thus, we naturally expect this relation to be broken for a non-metric spacetime, such as our area metric spacetime, possibly even in a way such that the result is no longer merely kinematical but dynamical, that is, dependent on the gravity theory. This turns out to be the case here, as we shall see in the following.

First of all, recall from Sec.~\ref{sec:light} that, at leading order in geometrical optics, area metric light propagation is essentially metric, so the standard ray optical derivation of the Etherington relation (e.g., Ref.~\refcite{sef92}, ch. 3) largely applies, yielding
\begin{equation}
D_L=\frac{(1+z)^2D_A}{\sqrt{1+\Delta}}\,,
\label{etherington}
\end{equation}
where $D_L, \ D_A, \ z$ are luminosity distance, angular diameter distance and redshift, respectively, as usual, and $\Delta$ is the photon excess fraction in a ray bundle domain $\mathcal{D}$ from the light source to the observer obeying
\begin{equation}
\Delta\cdot N=\int_{\mathcal{D}} \rmd^4x \sqrt{- \det P^\sharp}\  N^\mu{}_{;\mu}\,,
\label{delta}
\end{equation}
where the covariant derivative is with respect to metric $P^\sharp$ of (\ref{p}), $N$ is the photon number and $N^\mu$ the photon current, as before. In standard metric theory, photon conservation implies the vanishing of $N^\mu{}_{;\mu}$ in Eq. (\ref{delta}) whence $\Delta=0$ in Eq. (\ref{etherington}), and we recover the standard Etherington relation. However, in the area metric case, although conservation law (\ref{conservation}) applies, the $N^\mu{}_{;\mu}$ does not vanish whence $\Delta\neq 0$, and we find a modification of the Etheringon relation. Concretely, for linearized area metric Schwarzschild of Sec.~\ref{sec:gravity}, Yukawa-type corrections emerge,
\[
D_L=(1+z)^2 D_A\left(1+\frac{3\kappa M}{8\pi}\left(\frac{e^{-\mu r_s}}{r_s}-\frac{e^{-\mu r_o}}{r_o}\right)\right)\,,
\]
with Euclidean distances $r_o, r_s$ between the mass $M$ and the observer and light source, respectively. Thus, although at the optical geometry (and hence image positions and time delays) is the same as for metric Schwarzschild at leading order in $M$, the area metric volume measure implies a difference in the image magnification.

\section{Concluding remarks}
The constructive gravity approach allows the derivation of predictive gravitational dynamics from bihyperbolic kinematics. This enables the study of spacetimes beyond metric geometry without additional assumptions about the modified gravity theory. Thus, qualitatively new gravitational lensing effects due to non-metricities can be derived to provide possible observational tests. The first concrete example of this was described here, resulting in a modified Etherington relation for the linearized area metric Schwarzschild spacetime. The interested reader is also referred to session AT5 {\it Constructive Gravity} of this Marcel Grossmann Meeting.

\end{document}